\documentclass[conference]{IEEEtran}
\usepackage[letterpaper, left=1.02in, right=1.02in, bottom=1.01in, top=0.75in]{geometry}
\usepackage{cite}
\usepackage{amsmath,amssymb,amsfonts}
\usepackage[pdftex]{graphicx}
\usepackage[pdftex]{color}
\usepackage{caption,subcaption}
\usepackage{array, booktabs, multirow}
\newcommand{\PreserveBackslash}[1]{\let\temp=\\#1\let\\=\temp}
\usepackage{algorithm}
\usepackage[noend]{algpseudocode}

\usepackage{comment}

\begin{document}

\title{{Stochastic Image Transmission with CoAP for Extreme Environments}
\thanks{A part of this work was supported by JSPS KAKENHI Grant Number JP21H03399 and JST, ACT-I Grant Number JPMJPR18UL and Presto Grant Number JPMJPR2137, Japan.}
}

\author{
\IEEEauthorblockN{1\textsuperscript{st} Erina Takeshita}
\IEEEauthorblockN{2\textsuperscript{nd} Asahi Sakaguchi}
\IEEEauthorblockN{7\textsuperscript{th} Yu Nakayama}
\IEEEauthorblockA{
\textit{Institute of Engineering} \\
\textit{Tokyo University of Agriculture and Technology}\\
Tokyo, Japan \\
\{erina.takeshita, asahi.s, yu.nakayama\}@ieee.org}
\\
\IEEEauthorblockN{5\textsuperscript{th} Kazuki Maruta}
\IEEEauthorblockA{
\textit{Academy for Super Smart Society}\\
\textit{Tokyo Institute of Technology}\\
Tokyo, Japan\\
kazuki.maruta@ieee.org}
\and
\IEEEauthorblockN{3\textsuperscript{rd} Daisuke Hisano}
\IEEEauthorblockN{4\textsuperscript{th} Yoshiaki Inoue}
\IEEEauthorblockA{
\textit{Graduate School of Engineering}\\
\textit{Osaka University}\\
Osaka, Japan\\
\{hisano, yoshiaki\}@comm.eng.osaka-u.ac.jp}
\\
\\
\IEEEauthorblockN{6\textsuperscript{th} Yuko Hara-Azumi}
\IEEEauthorblockA{
\textit{School of Engineering}\\
\textit{Tokyo Institute of Technology}\\
Tokyo, Japan\\
hara@cad.ict.e.titech.ac.jp}
}

\maketitle

\begin{abstract}
% -200
Communication in extreme environments is an important research topic for various use cases including environmental monitoring.
A typical example is underwater acoustic communication for 6G mobile networks.
The major challenges in such environments are extremely high-latency and high-error rate.
They make real-time image transmission difficult using existing communication protocols.
This is partly because frequent retransmission in noisy networks increases latency and leads to serious deterioration of real-timeness.
To address this problem, this paper proposes a stochastic image transmission with Constrained Application Protocol (CoAP) for extreme environments.
The goal of the proposed idea is to achieve approximate real-time image transmission without retransmission using CoAP over UDP.
To this end, an image is divided into blocks, and value is assigned for each block based on the requirement.
By the stochastic transmission of blocks, the reception probability is guaranteed without retransmission even when packets are lost in networks.
We implemented the proposed scheme using Raspberry Pi 4 to demonstrate the feasibility.
The performance of the proposed image transmission was confirmed from the experimental results.
\end{abstract}


%%%%%%%%%%%%%%%%%%
%
% Introduction
%
%%%%%%%%%%%%%%%%%%
\section{Introduction}
% Extreme
Communication in extreme environments is an essential technology for achieving ubiquitous connectivity all over the world.
A typical example of extreme environments is underwater where the difficulty lies in the time-varying channel state and high-attenuation.
It is one of the major goals of 6G to integrate underwater networks to form a space-air-ground-underwater network~\cite{zhang20196g, dang2020should}.
Extreme communication will be an important platform for various applications including environmental monitoring.
The challenges of extreme communication are high-latency and high-error rate due to noisy and time-varying channels.

% Image transmission
Real-time object detection with deep learning (DL)~\cite{krizhevsky2012imagenet} is an essential technology for various monitoring applications.
Numerous network cameras are deployed for monitoring purposes and edge computing is a popular paradigm to leverage resource-limited devices.
The high-latency and high-error rate in extreme environments make real-time image transmission difficult using existing communication protocols.
To ensure reliability of data transfer, error correction coding such as FEC (Forward Error Correction) and retransmission schemes, e.g. Hybrid ARQ (Automatic Repeat reQuest) and fast retransmission in TCP, are popularly employed in communication systems.
However, frequent retransmission seriously increases latency in noisy networks so that the real-timeness of data transmission is deteriorated.

% this paper
Therefore, this paper proposes a stochastic image transmission with CoAP for extreme environments.
CoAP, which is defined in RFC7252, is a light-weight and asynchronous protocol based on UDP for resource-limited IoT devices.
The goal of the proposed scheme is to achieve approximate real-time image transmission without retransmission.
An image is divided into blocks, and the blocks are stochastically transmitted to guarantee the reception probability without retransmission even when packets are lost in noisy networks.
The rest of the paper is organized as follows.
The related works are described in section \ref{sec:rltd}.
Section \ref{sec:prp} introduces the proposed data transmission scheme.
%The performance of the proposed scheme is theoretically analyzed in section \ref{sec:theo}.
Section \ref{sec:prtcl} explains the protocol design for implementation of the proposed scheme.
Section \ref{sec:exp} reports the results of experimental validation.
The performance of the proposed scheme is demonstrated in section \ref{sec:prm}.
Finally, section \ref{sec:cncl} describes the concluding remarks.

%%%%%%%%%%%%%%%%%%
%
% Related work
%
%%%%%%%%%%%%%%%%%%
\section{Related work} \label{sec:rltd}
% retransmission
There have been a wide variety of approaches for ensuring data integrity in packet transmission.
Hybrid ARQ is a common protocol for error-correction in wireless networks, which is a combination of high-rate FEC and ARQ error control.
The original data is encoded with FEC, and the parity bits are either immediately sent with the packet or transmitted on request from a receiver that detects an erroneous packet.
Although the study on the performance of hybrid ARQ has a long history~\cite{caire2001throughput, zhao2005practical}, it still has been investigated with various mechanisms in the recent past, e.g. non-orthogonal multiple access (NOMA)~\cite{cai2018performance}, average age of information~\cite{ceran2019average}, and timely channel coding~\cite{arafa2019timely}.
Another typical mechanism for ensuring errorless data transfer is retransmission in TCP.
The optimum retransmission sequence has been investigated for various network environments such as wireless LANs~\cite{shin2016tcp}, data center networks~\cite{hwang2016fast}, and 5G mmWave networks~\cite{polese2017tcp}.

% image encoding
Image encoding and transfer protocols have been intensely studied over the past decades to reduce data size and latency without retransmission~\cite{turner1992image, danskin1995fast, wah2000survey}.
Priority encoding transmission (PET) was introduced for sending hierarchically organized messages over lossy networks~\cite{albanese1996priority, boucheron2000priority}.
A burst-erasure correcting code for low-delay transmission was presented in \cite{adler2017burst}.
The objective of this work is to minimize the average delay across the erased packets in a burst.
\cite{vo2017optimal} proposed a mobile video streaming scheme for 5G device-to-device (D2D) communication.
The proposed idea allocated optimal encoding rates to different layers of a video segment to packetize the segment into multiple descriptions with embedded forward error correction for improving quality of experience (QoE) of users.
A expanding-window batched sparse code was proposed in \cite{xu2017expanding} for scalable video multicasting over erasure networks with heterogeneous video quality requirements.
In this scheme, the input symbols are grouped into overlapped windows according to their importance levels so that the more important symbols are encoded with lower rate to be decoded by more destinations.

% contribution
The contribution of this paper is to propose an image transmission scheme for extreme environments that guarantees the arrival rate of certain regions of the image.
The proposed approach allows erroneous data transfer in networks without retransmission.
It achieves low-latency data reception that is sufficient for the DL-based real-time object recognition.
To the best of our knowledge, there has never been a similar approach in the recent literature.

%%%%%%%%%%%%%%%%%%
%
% Proposal
%
%%%%%%%%%%%%%%%%%%
\section{Stochastic image transmission} \label{sec:prp}
\subsection{Concept} \label{sec:prp_cncpt}
\begin{figure}[!t]
\centering
	\includegraphics[width=2.0in]{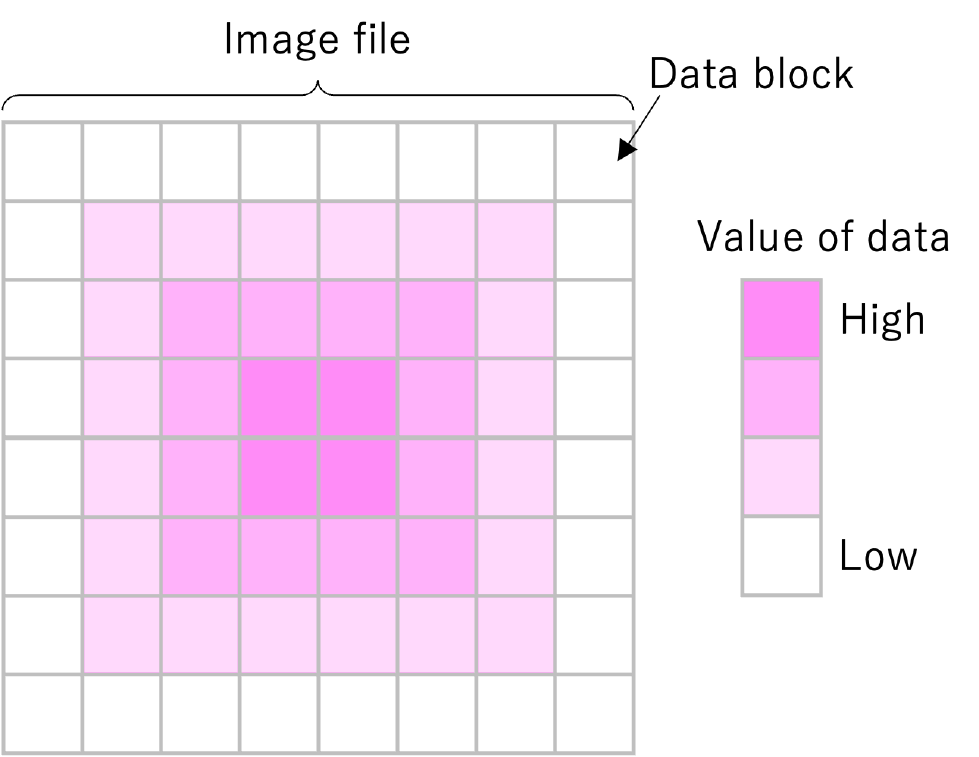}
	\caption{Value setting of data blocks in image.}
	\label{fig:datavalue}
\end{figure}

\begin{figure}[!t]
\centering
	\includegraphics[width=3.0in]{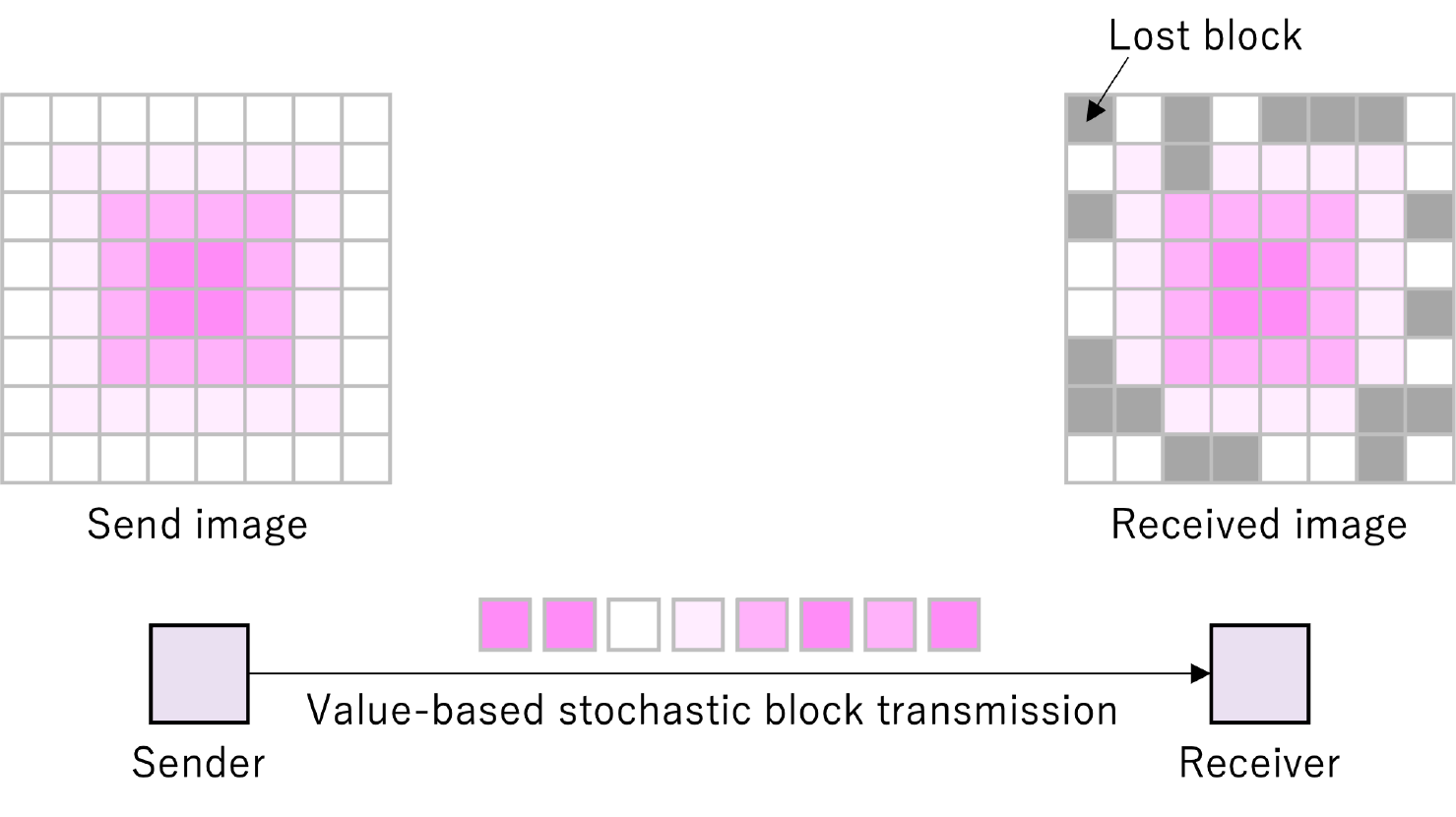}
	\caption{Value-based stochastic image transmission.}
	\label{fig:concept}
\end{figure}

This section introduces the proposed value-based stochastic image transmission for extreme environments.
The goal of the proposed scheme is to achieve approximate data transmission without retransmission for real-time object detection over networks.
The basic idea is to divide an image file into data blocks, and then assign value for each block as depicted in Fig.~\ref{fig:datavalue}.
The concept of the stochastic image transmission is shown in Fig.~\ref{fig:concept}.
The blocks are stochastically sent without retransmission to reduce latency.
The high-valued blocks are sent with higher probability to ensure the required reception probability in noisy networks.

\begin{figure}[!t]
\centering
	\includegraphics[width=3.0in]{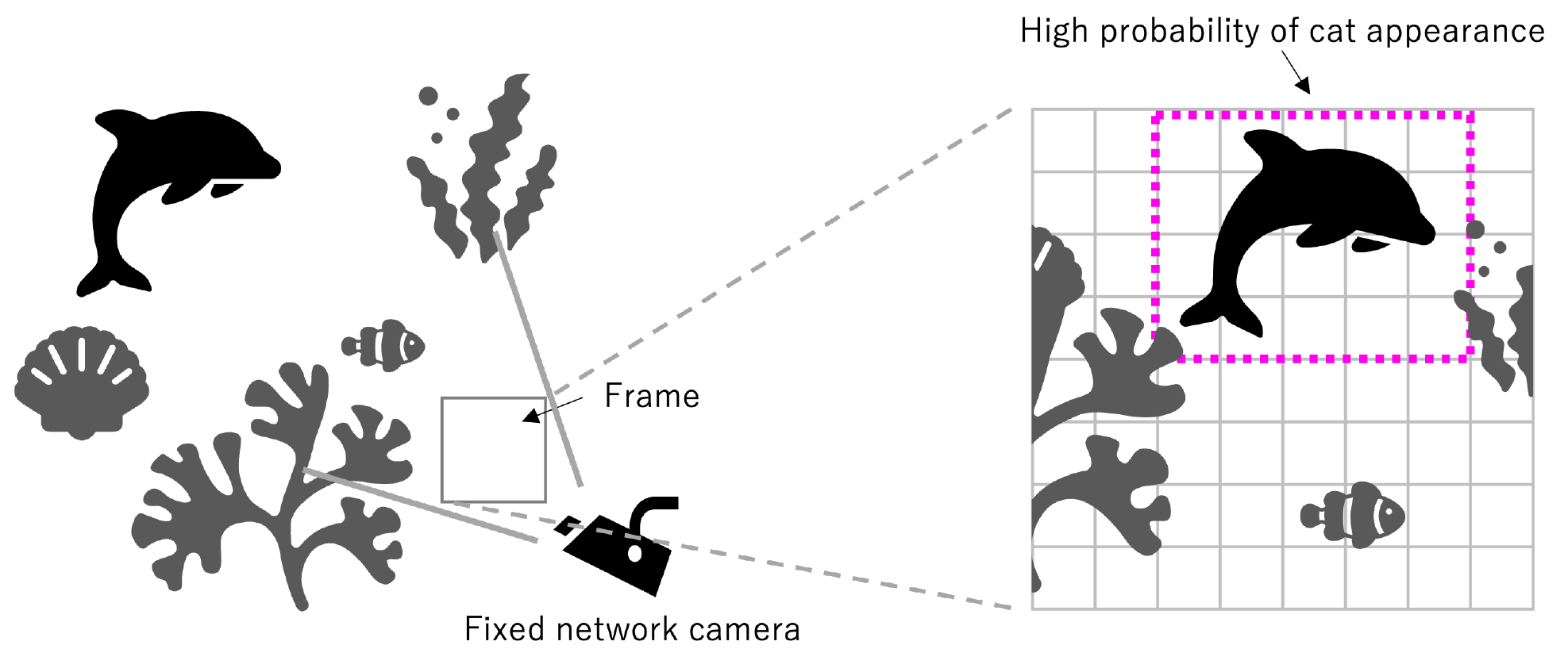}
	\caption{Example scenario for fixed network camera.}
	\label{fig:frame}
\end{figure}

The proposed scheme is based on the assumption that the value of an image data is not uniform.
That is, a target object to detect often appears at certain areas in the frame of a camera.
Let us introduce typical examples: the area where a target object passes is often determined by the environments where a fixed network camera is located (Fig.~\ref{fig:frame}).
Thus, certain areas in the frame can be defined as high-valued areas.
For the purpose of real-time object detection over networks, it is successful with a high probability if high-valued areas are received.
In other words, the purpose is achieved as long as the edge server can detect the object even if many packets are lost in a noisy network.

To achieve this, the proposed scheme stochastically sends data blocks in accordance with the defined value.
The advantage of the proposed idea is low-latency with no retransmission unlike TCP.
The redundancy for high-valued blocks is stochastically ensured, and thus the reception probability is guaranteed without retransmission even when packets are lost in lossy networks.
Furthermore, it can also contribute for reduction in the amount of transferred data depending on the parameter settings.

\subsection{Variable definition} \label{sec:prp_var}
The variables used in the proposed scheme are summarized in Table~\ref{tbl:variables}.
The detail of each variable is explained in the following sections.

\let\PBS=\PreserveBackslash
\begin{table}[!t]
	\renewcommand{\arraystretch}{1.2}
	\caption{Variables}
	\label{tbl:variables}
	\centering
	\begin{tabular}[t]{>{\PBS\centering\hspace{0pt}}p{0.5in} >{\PBS\centering\hspace{0pt}}p{2.25in}} \toprule
		Variable & Definition \\ \hline
		$\mathcal{I}$ & Set of data blocks \\
		$i$ & Data block identifier in $\mathcal{I}$ \\

		$s_{blk}$ & Data size of a block \\
		$s_{pkt}$ & Packet size \\
		$S_{org}$ & Original size of image \\
		$S_{prp}$ & Total transmission size of image \\

		$v_{i}$ & Value of $i$th block \\
		$p_{i}$ & Transmission probability of $i$th block \\

% 		$c_{org}$ & Packet count for original image \\
% 		$c_{prp}$ & Packet count with proposed scheme \\
		$e_{i}$ & Expected transmission count for $i$th block \\

		$L_{pkt}$ & Packet loss rate \\
		$r_{blk}$ & Transmission success probability for block  \\
		$\rho_{i}$ & Expected successful arrival count for $i$th block \\
		$R_{i}$ & Required reception probability \\

%		$$ &  \\
		\bottomrule
	\end{tabular}
\end{table}
\renewcommand{\arraystretch}{1}

\subsection{Data transmission} \label{sec:prp_trns}
The set of blocks that compose an image file is defined as $\mathcal{I}$.
Let $i \in \mathcal{I}$ denote the identifier for data block in an image file.
The value and the transmission probability of $i$th block are defined as $v_{i}$ and $p_{i}$, respectively.
The transmission probability is computed in accordance with the value, which is formulated as
\begin{equation}
p_{i} = \frac{v_{i}}{\sum_{i \in \mathcal{I}} v_{i}}.
\label{eq:trns_p}
\end{equation}
From the definition, the total transmission probability satisfies
\begin{equation}
\sum_{i \in \mathcal{I}} p_{i} = 1.
\label{eq:trns_total_p}
\end{equation}

The original data size of the image can be described as
\begin{equation}
S_{org} = |\mathcal{I}| s_{blk},
\label{eq:trns_s_org}
\end{equation}
where $s_{blk}$ is the size of a data block assuming that the size of each block is the same.
The packet count for transmitting a data block is described as
\begin{equation}
K = \frac{s_{blk}}{s_{pkt}},
\label{eq:trns_k}
\end{equation}
where $s_{pkt}$ denotes the packet size in the network.

The total transmission size of the image with the proposed scheme is described as $S_{prp}$, which is determined considering parameters such as the original data size $S_{org}$, error rate, and link capacity.
The total transmission count of data blocks is determined as
\begin{equation}
N = |\mathcal{I}| \frac{S_{prp}}{S_{org}},
\label{eq:trns_N}
\end{equation}
Thus, the expected transmission count for $i$th block is computed as
\begin{align}
e_{i} = p_{i} N.
\label{eq:trns_e}
\end{align}

\subsection{Data reception} \label{sec:prp_rcv}
The transmission success probability for a data block is formulated as
\begin{equation}
r_{blk} = 1 - L_{pkt}^K,
\label{eq:rcv_r_blk}
\end{equation}
where $L_{pkt}$ denote the packet loss rate in the network.
This is the probability that the receiver correctly receives the block without an error.

The non-arrival probability of $i$th block is formulated as
\begin{equation}
L_{i} = \{ (1 - p_{i}) + p_{i} L_{pkt}^K \}^N.
\label{eq:rcv_rho_non}
\end{equation}
Equation \eqref{eq:rcv_rho_non} represents the total probability of no-transmission and failed transmission with $N$ trials.
Thus, the arrival probability of $i$th block is
\begin{equation}
\rho_{i} = 1 - L_{i}.
\label{eq:rcv_rho}
\end{equation}

The requirement for data reception is defined as
\begin{equation}
\rho_{i} \geq R_{i},
\label{eq:rcv_rho2}
\end{equation}
where $R_{i}$ is the required reception probability for $i$th block.
The reception probability is predefined based on the characteristics of the image.

\subsection{Value setting} \label{sec:prp_val}
This section introduces the value setting to satisfy the requirement for data reception formulated in \eqref{eq:rcv_rho2}.
With \eqref{eq:rcv_rho_non} and \eqref{eq:rcv_rho}, the requirement in \eqref{eq:rcv_rho2} is written as
\begin{equation}
1 - \{ (1 - p_{i}) + p_{i} L_{pkt}^K \}^N \geq R_{i}.
\label{eq:val_req}
\end{equation}
This is transformed to be a constraint for $p_{i}$ as
\begin{equation}
p_{i} \geq \frac{1 - (1 - R_{i})^\frac{1}{N}}{1 - L_{pkt}^K}.
\label{eq:val_p}
\end{equation}
From \eqref{eq:trns_p}, the requirement for $i$th block becomes
\begin{equation}
\frac{v_{i}}{\sum_{i \in \mathcal{I}} v_{i}} \geq \frac{1 - (1 - R_{i})^\frac{1}{N}}{1 - L_{pkt}^K}.
\label{eq:val_req2}
\end{equation}

By totaling \eqref{eq:val_req2} for all $i \in \mathcal{I}$, we have
\begin{equation}
1 - L_{pkt}^K \geq |\mathcal{I}| - \sum_{i \in \mathcal{I}} (1 - R_{i})^\frac{1}{N}.
\label{eq:val_req_total}
\end{equation}
Therefore, the constraint for reception probability is formulated as
\begin{equation}
\sum_{i \in \mathcal{I}} (1 - R_{i})^\frac{1}{N} \geq |\mathcal{I}| - 1 + L_{pkt}^K
\label{eq:val_rcv_pr}
\end{equation}
This inequality is the constraint for setting the reception probability.
The loss rate $L_{pkt}$, the packet size $s_{pkt}$, and the original data size $S_{org}$ are determined by the network environments.
Thus, the range of $R_{i}$ is limited by the block size $s_{blk}$ and the total transmission size $S_{prp}$.
Assuming that \eqref{eq:val_rcv_pr} is satisfied, $v_{i}$ is set to satisfy equality in \eqref{eq:val_req}.

%%%%%%%%%%%%%%%%%%
%
% Protocol design
%
%%%%%%%%%%%%%%%%%%
\section{Protocol design} \label{sec:prtcl}
\subsection{Overview} \label{sec:prtcl_ovw}
This section introduces the protocol design for implementing the proposed stochastic image transmission scheme.
We employed CoAP which is defined in RFC 7252 as an asynchronous messaging protocol for resource-limited devices.
The advantages of CoAP are its simpleness, light-overhead, and the flexibility for confirmable/non-confirmable transmission.
CoAP assumes lossy and low-bandwidth networks so that it is suitable for extreme environments.
The stochastic transmission is implemented with the non-confirmable transmission of CoAP.
The blocks are identified with a newly defined \textit{Block Transmission header} (BT-header).
The detail is explained in the following.

\subsection{Packet format} \label{sec:prtcl_pkt}
\begin{figure}[!t]
\centering
	\includegraphics[width=2.5in]{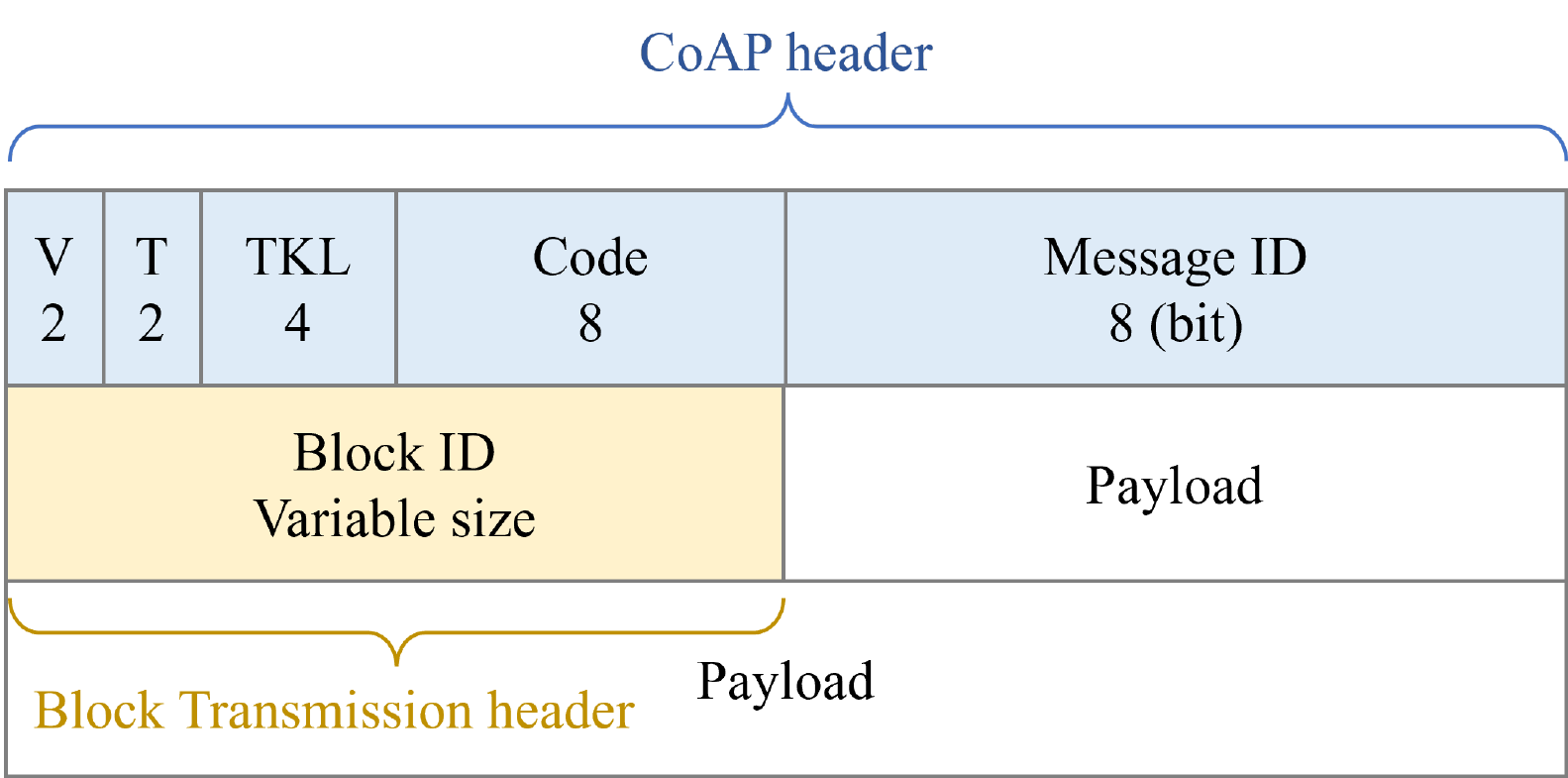}
	\caption{Packet format.}
	\label{fig:pkt_format}
\end{figure}

Fig.~\ref{fig:pkt_format} shows the packet format.
Since the numbers of blocks are variable in accordance with the image size, we define a BT-header as a variable length field following the CoAP header for identifying the block ID.
The Token and Options fields in the CoAP header are not used to reduce the packet size.
The BT-header consists of a variable length \textit{Block ID} field.
The length of \textit{Block ID} field is agreed among the client and the server before the image transmission.
This value is computed as $\lceil \log_2 |\mathcal{I}| \rceil$, which is determined by the image size.

\subsection{Sequence} \label{sec:prtcl_seq}
\begin{figure}[!t]
\centering
	\includegraphics[width=3.0in]{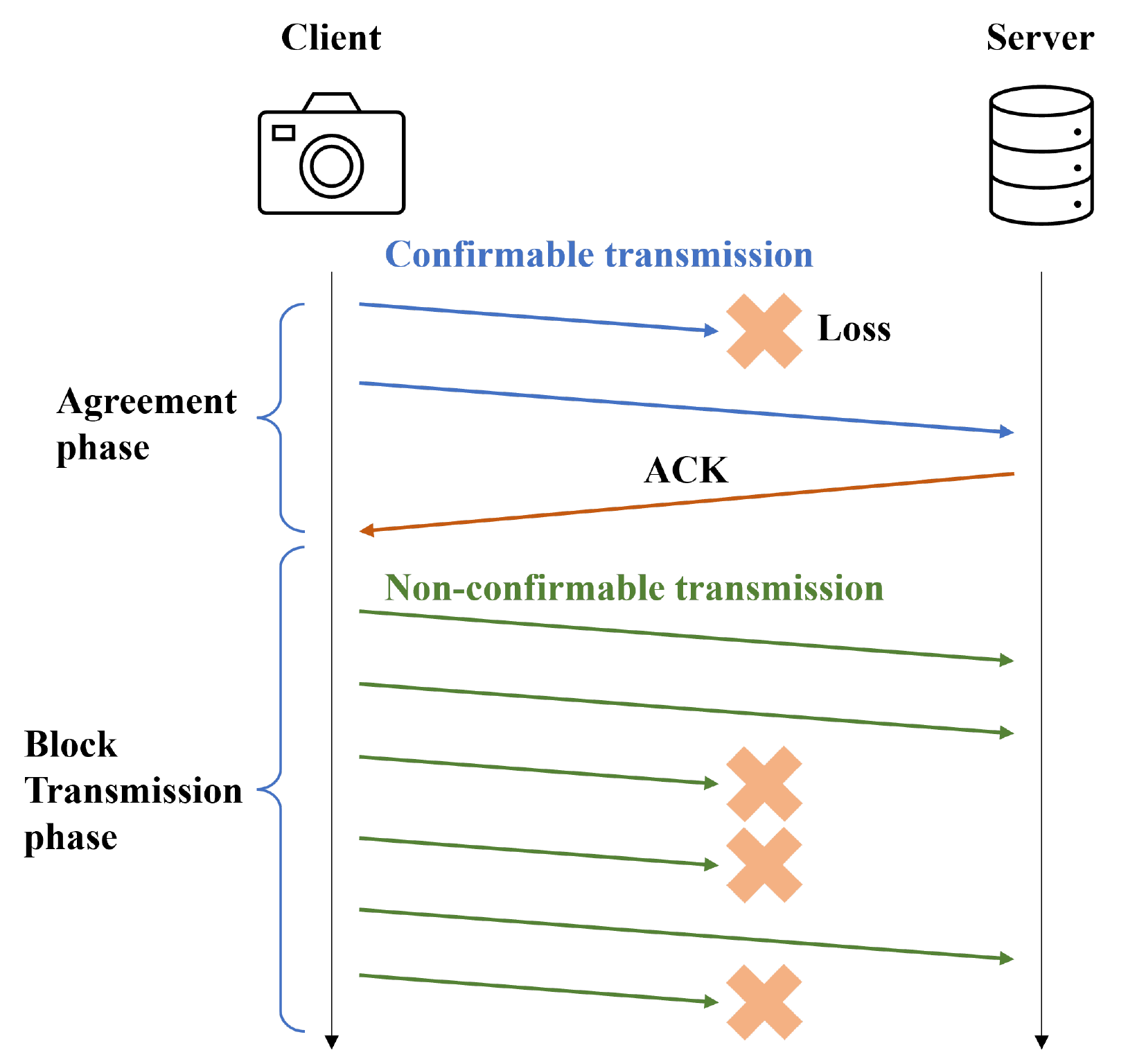}
	\caption{Sequence of proposed method.}
	\label{fig:exp_sequence}
\end{figure}

Fig.~\ref{fig:exp_sequence} shows the sequence of the image transmission among the client and the server.
The proposed sequence is twofold: the \textit{Agreement phase} and the \textit{Block transmission phase}.

\subsubsection{Agreement phase}
First, the client and the server agrees on the parameters of the stochastic image transmission.
The parameters are the length \textit{Block ID} field and the total transmission count $N$.
The transmission count $N$ is used for computing the expected transmission end time.
The client sends a request to the server with a confirmable message of CoAP to ensure the reliability in lossy networks.
The server sends an ACK message to confirm the end of the phase.

\subsubsection{Block transmission phase}
After the \textit{Agreement phase}, the stochastic image transmission is executed.
The client binarizes the original image file to data blocks.
It starts to stochastically send the data blocks.
They are forwarded with non-confirmable messages of CoAP to avoid retransmission.
After a certain time period, the server stops receiving data blocks to convert the received binary data to an image file.

%%%%%%%%%%%%%%%%%%
%
% Experimental validation
%
%%%%%%%%%%%%%%%%%%
\section{Experimental validation} \label{sec:exp}
This section introduces the experimental results to show the feasibility of the proposed scheme.

\subsection{Experimental Setup} \label{sec:exp_stup}
\subsubsection{Network topology} \label{sec:exp_stup_tp}
A 8GB Raspberry Pi 4 was employed as a client.
We used Python libraries \textit{aiocoap} and \textit{asyncio} to use CoAP.
The server was equipped with Intel Core i9-10980XE (3.0-4.6GHz, 18cores, 36threads) and NVIDIA GeForce RTX 3090.
The server OS was Ubuntu 20.04.
They were connected via 1Gbps Ethernet.
An extreme environment was simulated by stochastic packet loss.

\subsubsection{Dataset} \label{sec:exp_stup_dt}
A green ball was placed on top of a randomly moving robot.
%Fig.~\ref{exp:example_image} shows example images.
We shot total of 1000 images using a Panasonic HC-V360MS camera.
The image size was reduced to $256 \times 144$ pixels.
We trained a model to recognize the green ball using Darknet YOLOv3.

\begin{figure}[!t]
\centering
 \subcaptionbox{8 $\times$ 8 pixels/block \label{exp:heatmap8}}{\includegraphics[width=1.5in]{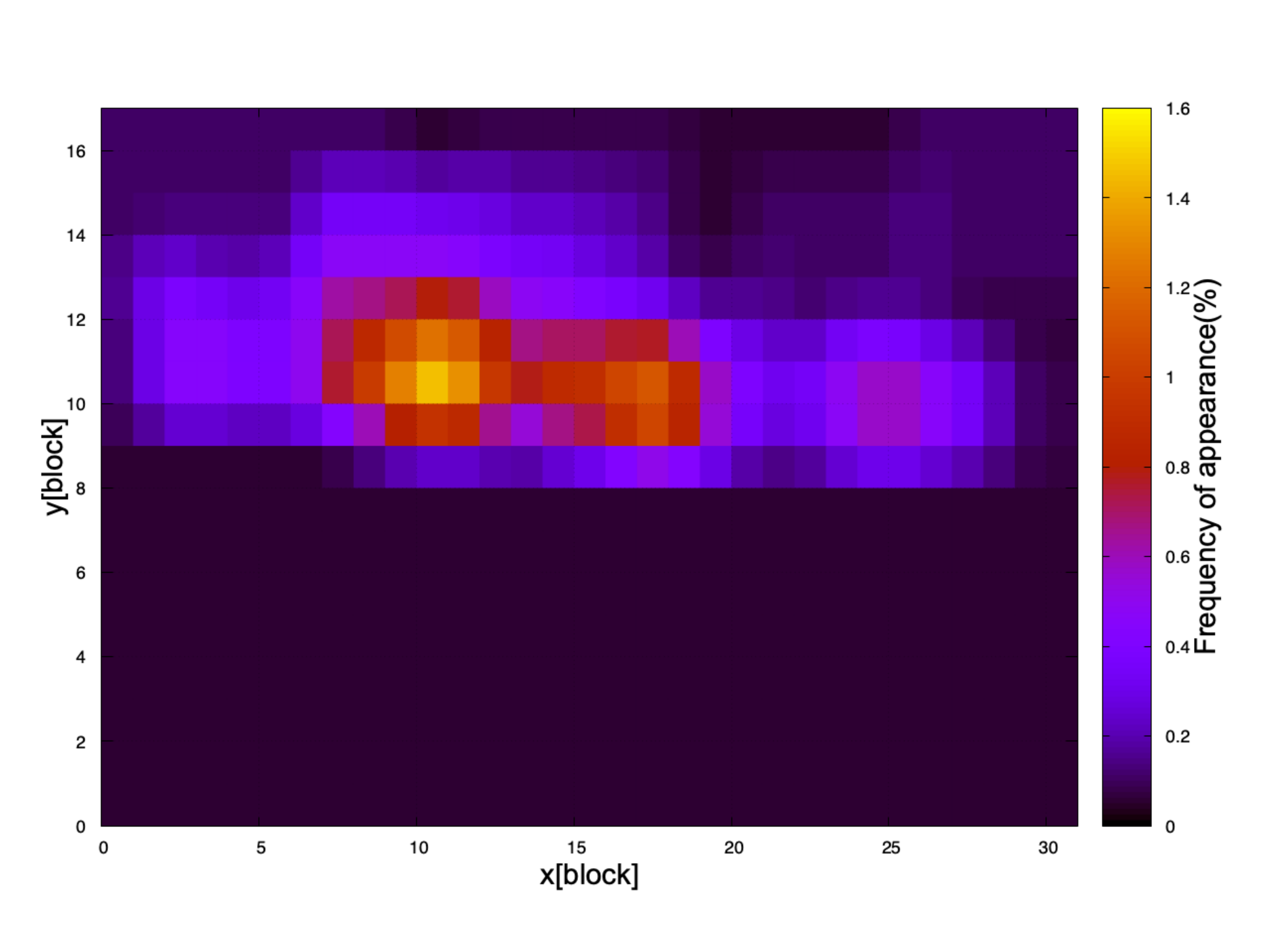}}
 \subcaptionbox{16 $\times$ 16 pixels/block \label{exp:heatmap16}}{\includegraphics[width=1.5in]{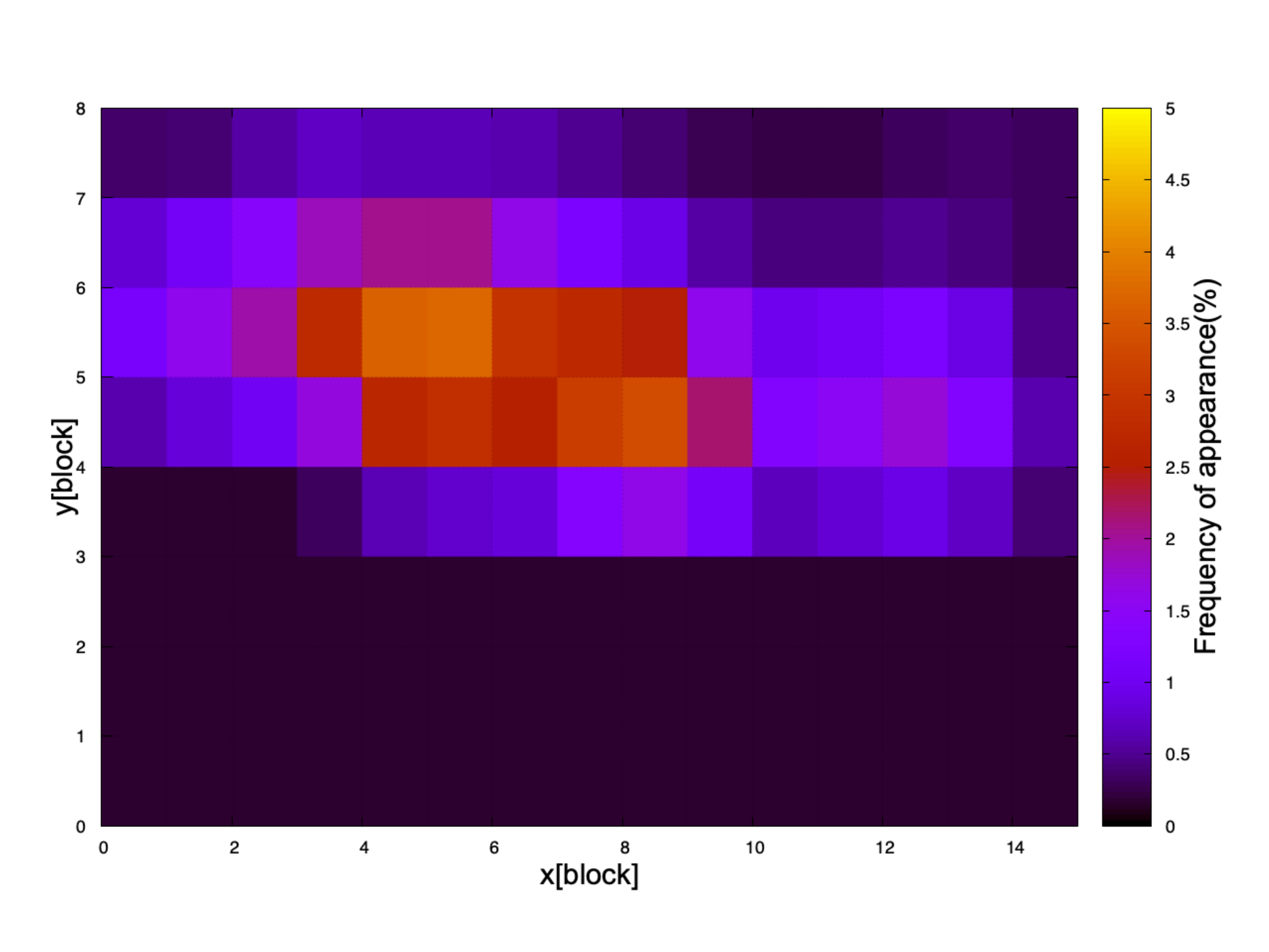}}
\caption{Distribution of target appearance frequency.}
\label{exp:heatmap}
\end{figure}

We employed two block sizes; $8 \times 8$ pixels and $16 \times 16$ pixels.
Fig.~\ref{exp:heatmap} shows the appearance frequency of the detected targets in each block.
The value of each block was set as a proportional distribution of the appearance frequency.

%\begin{figure}[!t]
%\centering
% \subcaptionbox{Ex. 1 \label{exp:data_1}}{\includegraphics[width=1.5in]{exp/data010.jpg}}
% \subcaptionbox{Ex. 2 \label{exp:data_2}}{\includegraphics[width=1.5in]{exp/data316.jpg}}
%\caption{Example images.}
%\label{exp:example_image}
%\end{figure}

\subsection{Results} \label{sec:exp_rslt}
\begin{figure}[!t]
\centering
 \subcaptionbox{8 $\times$ 8 pixels/block \label{exp:filling8}}{\includegraphics[width=1.5in]{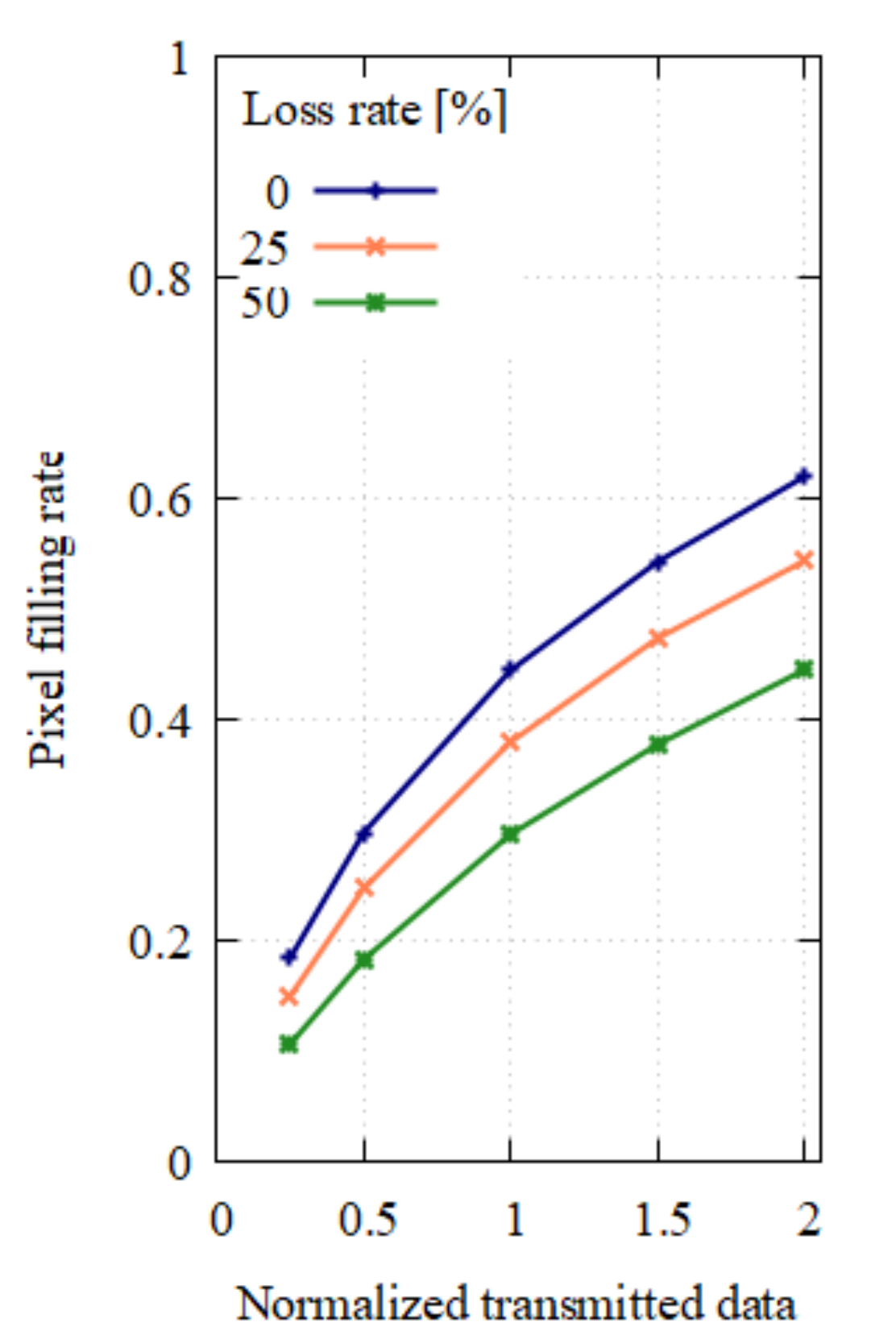}}
 \subcaptionbox{16 $\times$ 16 pixels/block \label{exp:filling16}}{\includegraphics[width=1.5in]{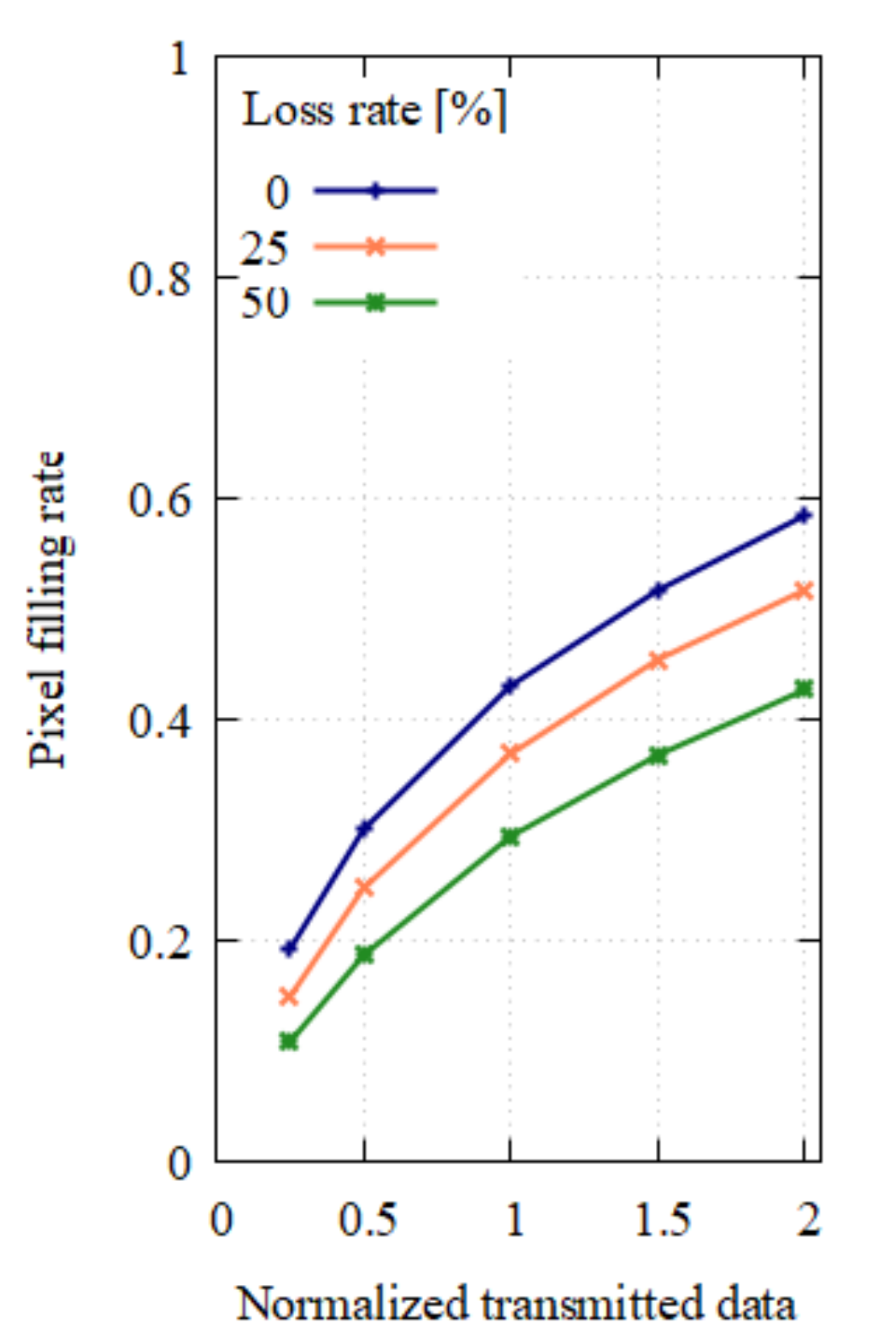}}
\caption{Pixel filling rate.}
\label{exp:filling}
\end{figure}

We defined a pixel filling rate as the ratio of received pixels to the original pixels.
The amount of transmitted data is normalized with the original image size.
Fig.~\ref{exp:filling} shows the pixel filling rate.
The pixel filling rate increases in accordance with the amount of transmitted data.
As the loss rate increases, the pixel filling rate decreases due to the loss of packets.
When the block size became large, the pixel filling rate decreased because of the overlap of received blocks.

\begin{figure}[!t]
\centering
 \subcaptionbox{8 $\times$ 8 pixels/block \label{exp:detection8}}{\includegraphics[width=1.5in]{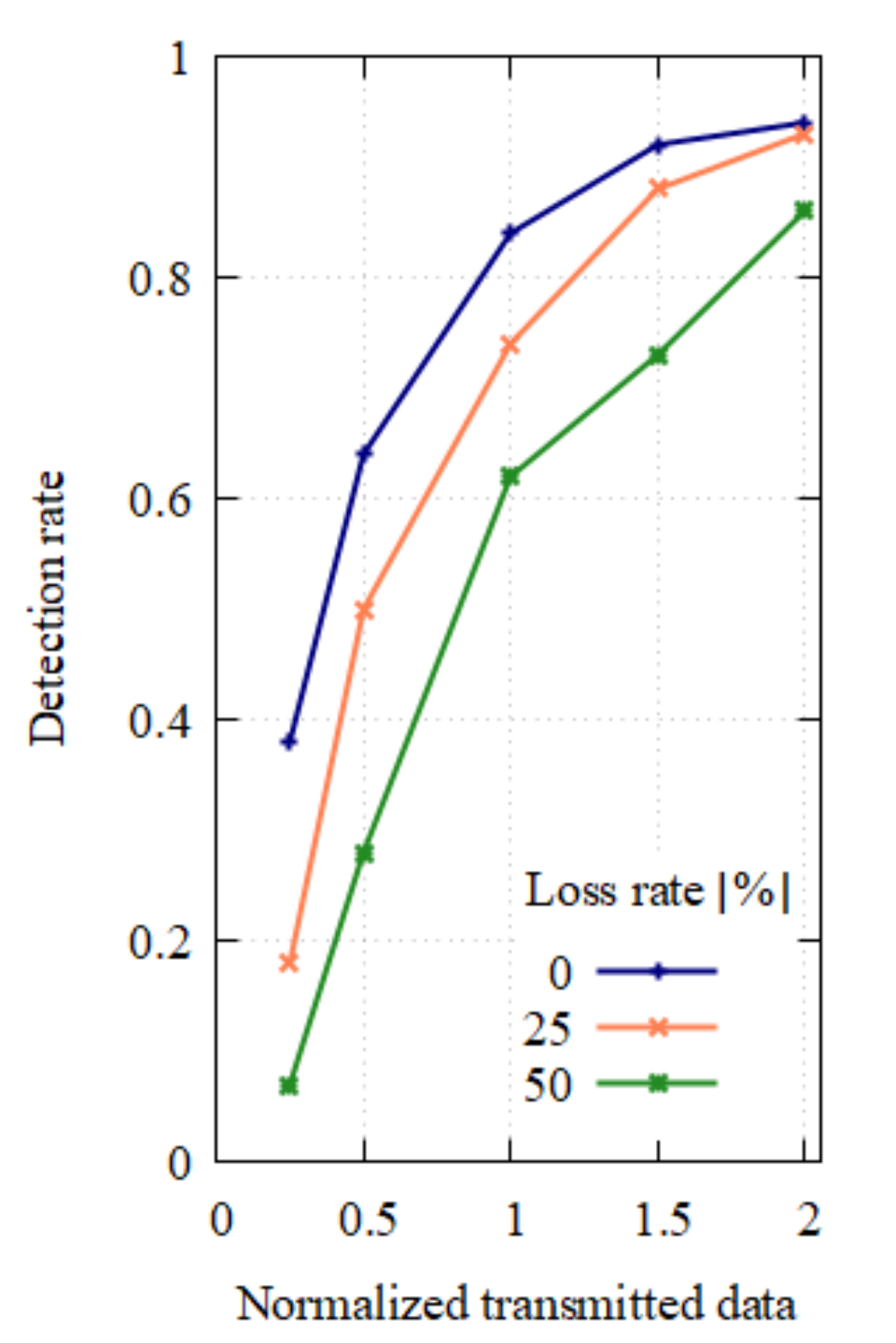}}
 \subcaptionbox{16 $\times$ 16 pixels/block \label{exp:detection16}}{\includegraphics[width=1.5in]{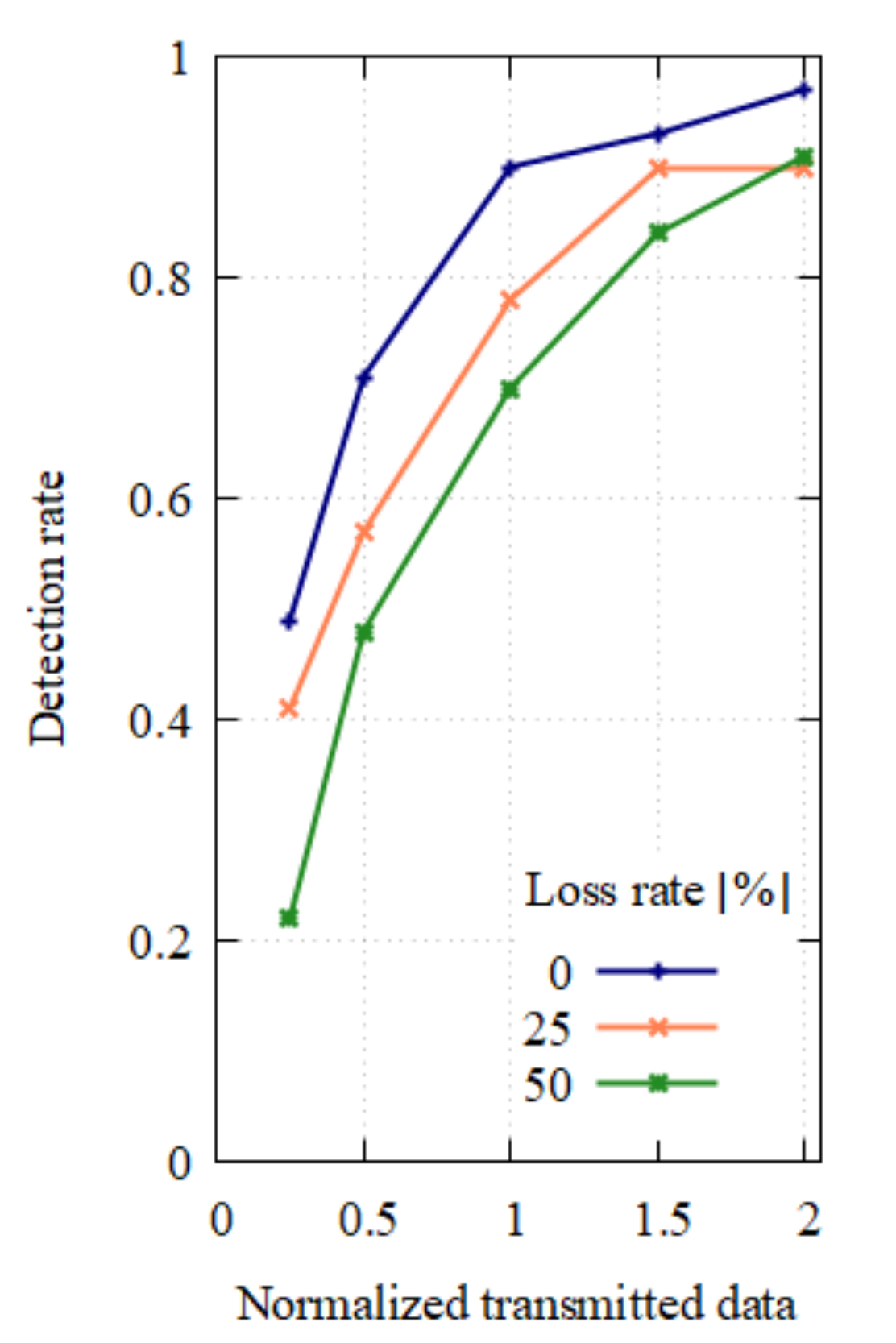}}
\caption{Detection rate.}
\label{exp:detection}
\end{figure}

We tried detecting the target from the received partial images using YOLOv3.
Fig.~\ref{exp:detection} shows the detection success rate.
The detection rate improved with the amount of transmitted data.
It was deteriorated by the packet loss.
Both block sizes achieved comparable performance.
When the normalized transmitted data was $1.0$, a detection rate of $70$\% was achieved even with a loss of $25$\ in both block sizes.
When the transmitted data was $2.0$, the detection rate was over $85$\% with $50$\% loss.

%Fig.~\ref{exp:relation} shows the relationship between the pixel filling rate and the detection rate.
%The detection rate became higher as the pixel filling rate increased.
%It exceeded $0.9$ with $50$\% of filled pixels.
%It was implied that valuable pixels were successfully transmitted with the proposed scheme even in extremely lossy networks.
These results show that the proposed stochastic image transmission over CoAP is suitable for a real-time monitoring purpose by exploiting the bias of image appearance probability.

%\begin{figure}[!t]
%\centering
% \subcaptionbox{8 $\times$ 8 pixels/block \label{exp:relation8}}{\includegraphics[width=1.5in]{graph/8compVSrecog.pdf}}
% \subcaptionbox{16 $\times$ 16 pixels/block \label{exp:relation16}}{\includegraphics[width=1.5in]{graph/16compVSrecog.pdf}}
%\caption{Pixel filling rate vs Detection rate.}
%\label{exp:relation}
%\end{figure}

%%%%%%%%%%%%%%%%%%
%
% Performance evaluation
%
%%%%%%%%%%%%%%%%%%
\section{Performance evaluation} \label{sec:prm}
\subsection{Setup} \label{sec:prm_stp}
The performance of the proposed scheme was evaluated in a lossy and low-bandwidth network.
An 8GB Raspberry Pi 4 was installed as an access point and a server.
Two 8GB Raspberry Pi 4 and a laptop were connected to the server as clients via IEEE802.11g.
They were placed on the other side of two walls to simulate an extreme environment.
The average RSSI was $-74.63$ dBm.
The packet loss rate was 2.839 \%.

Two client devices constantly sent UDP packets as background traffic.
The other client sent $100$ image files with the proposed scheme.
We measured the required time to detect the target object.
The proposed scheme was compared with TCP-based image transmission.
The congestion control algorithm was CUBIC.

\subsection{Result} \label{sec:prm_rslt}
\begin{figure}[!t]
\centering
	\includegraphics[width=3.0in]{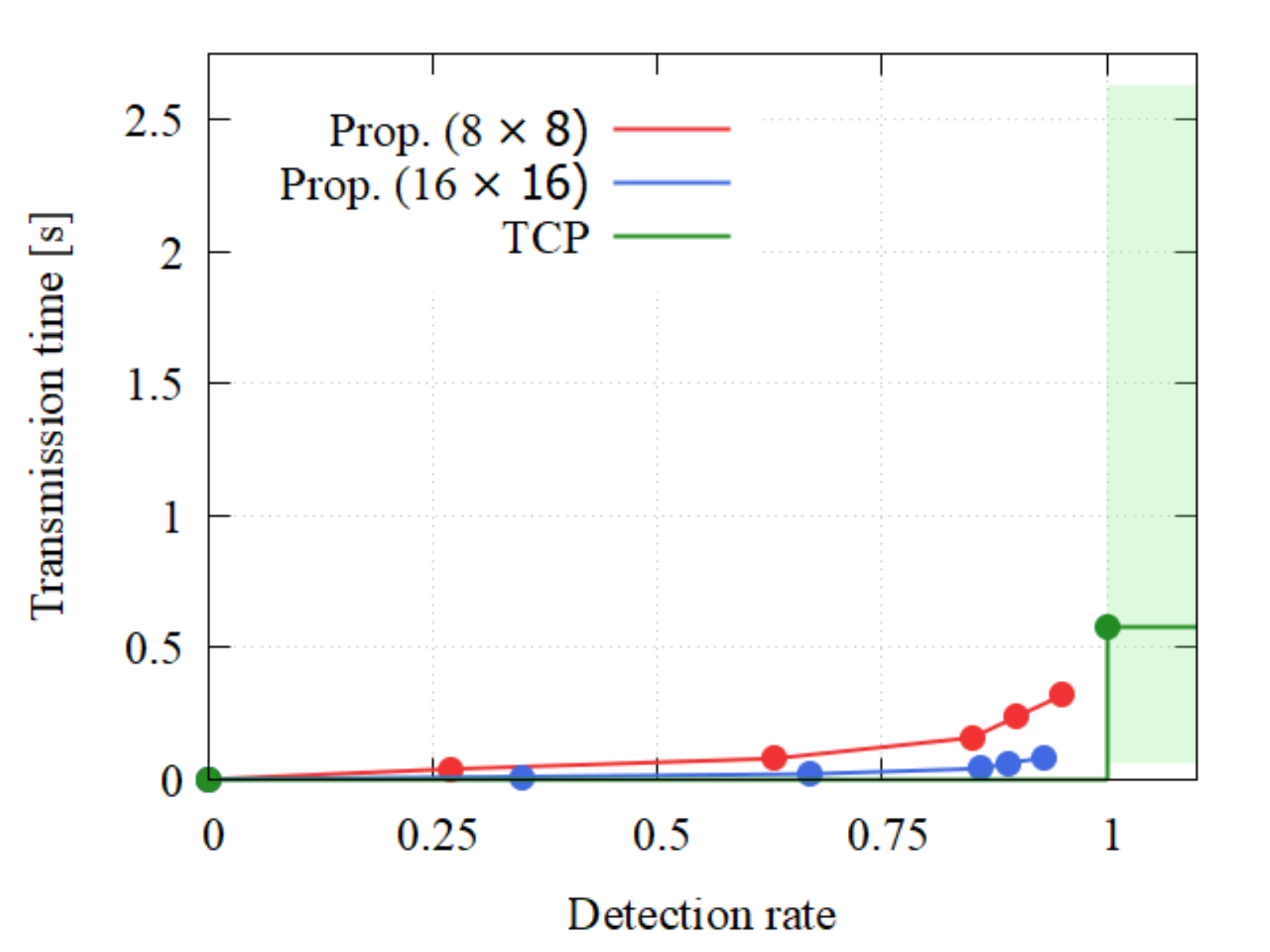}
	\caption{Detection rate vs transmission time.}
	\label{fig:prm_rslt}
\end{figure}

Fig.~\ref{fig:prm_rslt} shows the relationship between the detection rate and the average transmission time for an image.
The proposed scheme achieved over $90$\% detection rate within $0.3$ seconds.
The existing TCP-based transmission required average $0.58$ seconds for transmitting an image in the condition due to high packet loss rate.
The worst case was $2.63$ seconds with TCP.
From this result, we confirmed the advantage of the proposed scheme in extreme environments.

\subsection{Discussion} \label{sec:exp_dscs}
The experimental results demonstrated the feasibility of the proposed scheme using resource-limited devices.
Note that the performance of the object detection rate largely depends on the bias in the occurrence of targets.
In the experimental condition, the frequency distribution was strongly biased so that the blocks around the center of images were highly valuable.
It constitutes future work to perform more comprehensive experiments in noisy wireless networks such as underwater acoustic communications.

%%%%%%%%%%%%%%%%%%
%
% Conclusion
%
%%%%%%%%%%%%%%%%%%
\section{Conclusion} \label{sec:cncl}
This paper proposed the stochastic image transmission with CoAP for extreme environments.
The goal is to achieve approximate data reception without retransmission for real-time object recognition in extreme environments.
It leverages CoAP which is an asynchronous protocol for resource-limited IoT devices.
With the proposed scheme, an image data is divided into blocks.
The blocks are stochastically sent with CoAP over UDP in accordance with the value assigned to each block based on the condition.
The high-valued blocks are sent with higher probability to ensure the required reception probability in noisy networks.
The latency caused by retransmission is drastically reduced with the proposed idea.
The proposed scheme was implemented using Raspberry Pi 4.
The feasibility of the proposed idea was confirmed via the experimental results.
The future work is more comprehensive experiments in extreme environments using narrow wireless networks.

%%%%%%%%%%%%%%%%%%
%
% Acknowledgment
%
%%%%%%%%%%%%%%%%%%
\section*{Acknowledgment}
A part of this work was supported by JSPS KAKENHI Grant Number JP21H03399 and JST, ACT-I Grant Number JPMJPR18UL and Presto Grant Number JPMJPR2137, Japan.

%%%%%%%%%%%%%%%%%%
%
% Bibliography
%
%%%%%%%%%%%%%%%%%%
\bibliographystyle{IEEEtran.bst}
\bibliography{stochastic2022takeshita}

\end{document}